\documentclass{elsart1p}
\usepackage{graphicx}
\usepackage{amssymb}
\begin{document}
\begin{frontmatter}
%
%
%
\title{ Probing photon structure in DVCS on a photon target }
%
%

\author[ad1,ad2]{M. El Beiyad},
\author[ad3]{S. Friot},
\author[ad1]{B.Pire},
\author[ad4]{L. Szymanowski},
\author[ad2]{S. Wallon}
\address[ad1]{ CPhT, \'Ecole Polytechnique, CNRS, Palaiseau, France}
\address[ad2]{LPT, Universit\' e Paris XI, CNRS, Orsay, France}
\address[ad3]{IPN, CNRS, Orsay, France}
\address[ad4]{Soltan Institute for Nuclear Studies, Warsaw, Poland}

\begin{abstract}
The factorization of the amplitude for the deeply virtual Compton scattering (DVCS) process $\gamma^*(Q) \gamma \to \gamma \gamma$ at high $Q^2$ is demonstrated in two distinct kinematical domains, allowing to 
define the photon generalized parton distributions and the diphoton generalized distribution amplitudes. Both these quantities exhibit an anomalous scaling behaviour and obey new inhomogeneous QCD evolution equations.
\end{abstract}
\end{frontmatter}
%
\section{Motivation}
\label{}
The parton content of the photon has been the subject of many studies since the 
seminal paper by Witten \cite{Witten} which allowed to define the anomalous quark and gluon distribution
functions. Recent progresses in exclusive hard reactions  focus on generalized parton distributions (GPDs), which are defined as Fourier transforms 
of matrix elements between different states, such as $
\langle N'(p',s') |\bar\psi(-\lambda n)\gamma.n\psi(\lambda n) | N(p,s) \rangle$
and their crossed versions, the generalized distribution amplitudes (GDAs) which describe the exclusive hadronization of a $\bar q q $ or $g g$ pair 
in a pair of hadrons, see Fig.~1.
\begin{figure}
\hspace*{2.5cm}\includegraphics[width=9cm]{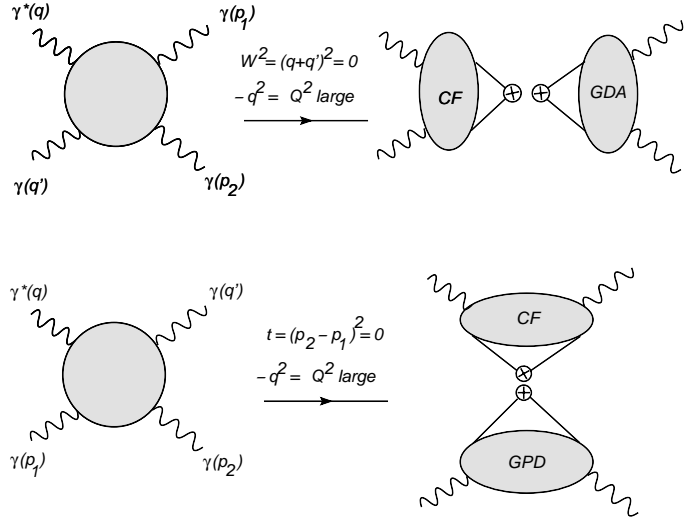}
\caption{Factorizations of the DVCS process on the photon.}
\end{figure}
 In the photon case, these quantities are perturbatively calculable \cite{Friot,El} at leading order in $\alpha_{em}$ and leading logarithmic order in $Q^2$. They constitute an interesting theoretical laboratory for the non-perturbative hadronic objects that hadronic GPDs and GDAs are.

\section{The diphoton generalized distribution amplitudes}
Defining the momenta as $q= p - \frac{Q^2n}{s} ,~q'= \frac{Q^2n}{s} ,~p_1=\zeta p ,~p_2= (1-\zeta) p$, where 
$p$ and $n$ are two light-cone Sudakov vectors and $2 p\cdot n =s $,
the amplitude of the process
\begin{equation}
\gamma^*(Q, \epsilon) \gamma(q', \epsilon') \to \gamma(p_{1}, \epsilon_1) \gamma(p_2,\epsilon_2)
\label{dvcs}
\nonumber
\end{equation}
may be written as
$A = \epsilon_\mu\epsilon'_\nu{\epsilon_1}^*_\alpha{\epsilon^*_2}_\beta T^{\mu\nu\alpha\beta}$.
In forward kinematics where $(q+q')^2=0$,
the tensorial decomposition of $T^{\mu\nu\alpha\beta}$ reads (see \cite{El})
\begin{eqnarray}
&&
\frac{1}{4}g^{\mu\nu}_Tg^{\alpha\beta}_T W_1^q+
\frac{1}{8}\left(g^{\mu\alpha}_Tg^{\nu\beta}_T 
+g^{\nu\alpha}_Tg^{\mu\beta}_T -g^{\mu\nu}_Tg^{\alpha\beta}_T \right)W_2^q
+ \frac{1}{4}\left(g^{\mu\alpha}_Tg^{\nu\beta}_T - g^{\mu\beta}_Tg^{\alpha\nu}_T\right)W_3^q\, .
\nonumber
\end{eqnarray}
At leading order, the three scalar functions $W_i^q$ can be written in a factorized form which is particularly simple when the factorization scale $M_F$ equals the photon virtuality $Q$. $W_1^q$ is then the convolution 
$W^q_{1}= \int\limits_{0}^1 dz \, C_V^q(z) \, \Phi_1^q(z,\zeta,0)$ of the coefficient function
$ C_{V}^q = e_q^2\left(\frac{1}{z}-\frac{1}{1-z}\right) $
with the {\em anomalous} vector GDA ($\bar z = 1-z, \bar \zeta = 1 - \zeta$) :
\begin{eqnarray}
\label{Phi1}
&&\Phi_1^q(z,\zeta,0) =
 \frac{N_C\,e_{q}^2}{2\pi^2} \log{\frac{Q^2}{m^2}}
\left[\frac{\bar{z}(2z-\zeta)}{\bar{\zeta}}\theta(z-\zeta)+\frac{\bar{z}(2z-\bar{\zeta})}{\zeta}\theta(z-\bar{\zeta}) \right. \nonumber \\
&&\hspace*{4cm}+\left.\frac{z(2z-1-\zeta)}{\zeta}\theta(\zeta-z)+\frac{z(2z-1-\bar{\zeta})}{\bar{\zeta}}\theta(\bar{\zeta}-z)\right] .
\nonumber
\end{eqnarray}
Conversely, $W_3^q$ is the convolution  of the  function $C_{A}^q = e_q^2\left(\frac{1}{z}+\frac{1}{\bar z} \right)$
with the axial GDA :
\begin{eqnarray}
\label{Phi3}
\hspace*{-0.5cm}\Phi_3^q(z,\zeta,0) = \frac{N_C\,e_{q}^2}{2\pi^2} \log{\frac{Q^2}{m^2}}\left[\frac{\bar{z}\zeta}{\bar{\zeta}}\theta(z-\zeta)-\frac{\bar{z}\bar{\zeta}}{\zeta}\theta(z-\bar{\zeta})
- 
\frac{z\bar{\zeta}}{\zeta}\theta(\zeta-z)+\frac{z\zeta}{\bar{\zeta}}\theta(\bar{\zeta}-z)\right] \;
\nonumber
\end{eqnarray} 
and $W^q_2=0$.
Note that these GDAs are not continuous at the points $z =\pm \zeta$. The anomalous nature of $\Phi^q_1$ and $\Phi^q_3$ comes from their proportionality to $\log{\frac{Q^2}{m^2}}$, which reminds us of the anomalous photon structure functions. A consequence is that 
$\frac{d}{d \,\ln Q^2}\Phi^q_i \neq 0$; consequently the QCD evolution equations of the diphoton GDAs obtained with the help of the ERBL kernel are non-homogeneous ones.

\section{The photon generalized parton distributions}
We now look at the same process in different kinematics, namely  $q=-2\xi p+n ,~q'=(1+\xi) p ,~p_1=n ,~p_2= p_{1} + \Delta = (1-\xi) p$~,
where $W^2= \frac{1-\xi}{2\xi}Q^2$ and $t=0$.
The tensor
$T^{\mu\nu\alpha\beta}$ is now decomposed on different tensors  with the help of three  functions ${\cal W}^q_i$ as (see \cite{Friot}):
\begin{eqnarray}
\frac{1}{4}g^{\mu\alpha}_Tg^{\nu\beta}_T {\cal W}^q_1
+
\frac{1}{8}\left(g^{\mu\nu}_Tg^{\alpha\beta}_T 
+g^{\alpha\nu}_Tg^{\mu\beta}_T -g^{\mu\alpha}_Tg^{\nu\beta}_T \right){\cal W}^q_2
+ \frac{1}{4}\left(g^{\mu\nu}_Tg^{\alpha\beta}_T - g^{\mu\beta}_Tg^{\nu\alpha}_T\right){\cal W}^q_3\, 
\nonumber
\end{eqnarray}
 These functions can also be written in factorized forms which have direct parton model interpretations when  the factorization scale $M_F$ is equal to $Q$ .
The coefficient functions are $C_{V/A}^q = - 2e_q^2\left(\frac{1}{x-\xi+i\eta} \pm \frac{1}{x+\xi-i\eta}\right)$ and the unpolarized $H_{1}^q$ and polarized $H_{3}^q$ anomalous GPDs of quarks inside a real photon read :
\begin{eqnarray}
\label{H1}
&&\hspace*{-1cm}H_{1}^q (x, \xi, 0) =
\frac{N_C\,e_{q}^2}{4\pi^2} \left[\theta(x-\xi)  \frac {x^2 + (1-x)^2-\xi^2}{1-\xi^2}\;\right. 
\nonumber\\
 &&\hspace*{1cm}
+ \left.\theta(\xi-x) \theta(\xi+x) \frac{x(1-\xi)}{\xi(1+\xi)} \;
- \theta(-x-\xi)  \frac{x^2+(1+x)^2-\xi^2}{1-\xi^2}\right]\; \ln\frac{Q^2}{m^2}\;,
\nonumber
\end{eqnarray}
\begin{eqnarray}
\label{H3}
&&\hspace*{-1cm}
H_{3}^q (x, \xi, 0) = \frac{N_C\,e_{q}^2}{4\pi^2} \left[\theta(x-\xi)   \frac {x^2 - (1-x)^2-\xi^2}{1-\xi^2}\;\right. \nonumber\\ &&\hspace*{1cm}
- \left.
\theta(\xi-x) \theta(\xi+x) \frac{1-\xi}{1+\xi} \;
+ \theta(-x-\xi)  \frac {x^2 - (1+x)^2-\xi^2}{1-\xi^2}\right]\; \ln\frac{Q^2}{m^2}\;.
\nonumber
\end{eqnarray}
Similarly as in the GDA case,  the anomalous generalized parton distributions $H_{i}^q$ are proportional to 
$\ln\frac{Q^2}{m^2}$. Consequently, the anomalous terms $H_{i}^q$ supply to the usual homogeneous DGLAP-ERBL evolution equations of GPDs a non-homogeneous term which changes  them into non-homogeneous evolution equations. 

We do not anticipate a rich phenomenology of these photon GPDs, but in the case of a high luminosity electron - photon collider which is not realistic in the near future. However, the fact that one gets explicit expressions for these GPDs may help to understand the meaning of general theorems such as the polynomiality and positivity \cite{pos} constrains or the analyticity structure \cite{ana}. For instance, one sees that a D-term in needed when expressing the photon GPDs in terms of a double distribution. One also finds that,  in the DGLAP region,  $H_1(x,\xi)$ is smaller than its positivity bound by a sizeable and slowly varying factor, which is of the order of $0.7 - 0.8$ for $\xi  \approx 0.3$.

\noindent
This work is supported by the Polish Grant N202 249235, the French-Polish scientific agreement Polonium
and the grant ANR-06-JCJC-0084.

\end{document}